\documentclass[11pt,a4paper]{article}

\usepackage{epsfig,amsmath,amssymb,cite}

\def\be{\begin{equation}}

\def\ee{\end{equation}}

\begin{document}

\title{General formalism for the stability of thin-shell wormholes\\
in 2+1 dimensions} 

\author{Cecilia Bejarano$^{1,}$\thanks{e-mail: cbejarano@iafe.uba.ar}, Ernesto F. Eiroa$^{1,2,}$\thanks{e-mail: eiroa@iafe.uba.ar}, 
Claudio Simeone$^{2,3,}$\thanks{e-mail: csimeone@df.uba.ar}\\
{\small $^1$ Instituto de Astronom\'{\i}a y F\'{\i}sica del Espacio, Casilla de correo 67, Sucursal 28, 1428,}\\
{\small Buenos Aires, Argentina}\\
{\small $^2$ Departamento de F\'{\i}sica, Facultad de Ciencias Exactas y Naturales,} \\ 
{\small Universidad de Buenos Aires, Ciudad Universitaria Pabell\'on I, 1428, 
Buenos Aires, Argentina}\\ 
{\small $^3$ IFIBA--CONICET, Ciudad Universitaria Pabell\'on I, 1428, 
Buenos Aires, Argentina}}

\maketitle

\begin{abstract}

In this article we theoretically construct circular thin-shell wormholes in a $2+1$ dimensional spacetime. The construction is symmetric with respect to the throat. We present a general formalism for the study of the mechanical stability under perturbations preserving the circular symmetry of the configurations, adopting a linearized equation of state for the exotic matter at the throat. We apply the formalism  to several examples.\\



\end{abstract}

\section{Introduction}

Traversable wormhole geometries \cite{motho,visser} have been widely studied in the last three decades. Several articles considering wormholes in low dimensional ($2+1$) spacetimes have appeared in the literature \cite{mth3d}. In the framework of  General Relativity, wormholes must be threaded by matter not satisfying the energy conditions.  A particular class of wormholes consists of those   constructed by cutting and pasting two manifolds to obtain a new one with a joining shell at the throat \cite{visser}. In recent years shells around vacuum (bubbles), around black holes and stars, and supporting traversable wormholes, have received considerable attention \cite{sta,stars,poivi,wh4g,wh4};  spherically symmetric shells have been studied in detail in four and also in more spacetime dimensions \cite{wh5,eisi12a}. We have recently analyzed four dimensional spherical shells for Einstein gravity coupled to Born--Infeld electrodynamics in relation with thin-shell wormholes \cite{risi} and shells around vacuum or around black holes \cite{eisi-bi}. The dynamical evolution of collapsing shells in three spacetime dimensions has been presented and applied to several examples in Refs. \cite{crisol}. Shells in a three dimensional background within Einstein--Maxwell theory have been associated to thin-shell wormholes \cite{raha}, and the analysis of $2+1-$dimensional charged shells around vacuum and around black holes in Born--Infeld electrodynamics was introduced in our paper \cite{eisi13}. Cylindrical thin-shell wormholes have also been studied, which because of their symmetry along their axis constitute a related problem; see for instance Refs. \cite{cyl}.

The most famous black hole solution in $2+1$ dimensions is the  Ba\~nados--Teitelboim--Zanelli (BTZ) geometry \cite{btz}. In the non rotating case the metric has the form
\be
ds^2=-f(r)dt^2+f^{-1}(r)dr^2+r^2d\theta^2,\label{1}
\ee
where 
\be
f(r)= -M-\Lambda r^2-2Q^2\ln \left( \frac{r}{r_0}\right) .\label{2}
\ee
The dimensionless constant $M$ is identified as the Arnowitt--Deser--Misner (ADM) mass, $Q$ is the electric charge, and $\Lambda=-l^{-2}$ is the cosmological constant with dimensions $[\rm{length}]^{-2}$.   The metric has the right signature only if $\Lambda <0$. So in what follows we will assume that  $\Lambda <0$. This choice makes possible a standard horizon structure in which the metric function is always positive beyond a certain radius, and the geometry is asymptotically Anti-de Sitter. This behavior has made this metric of great interest within the framework of string theory.

Another interesting $2+1$ dimensional geometry, obtained by Chan and Mann \cite{chanmann}, is associated to the existence of a dilaton field coupled to the electromagnetic field. In the presence of  a cosmological constant background, the spherically symmetric solution includes a dilaton with a radial logarithmic behavior, and a one-parameter family of metrics of the form
\be
ds^2=-f(r)dt^2+\frac{4r^{\frac{4}{N}-2}dr^2}{N^2\gamma^{\frac{4}{N}}f(r)}+r^2d\theta^2,
\ee
with
\be
f(r)=-\frac{2M}{N}r^{\frac{2}{N}-1}-\frac{8\Lambda r^2}{(3N-2)N}+\frac{8Q^2}{(2-N)N}.\label{cm}
\ee
Here $M$ is the mass, $Q$ is the electric charge, $\Lambda$ is the cosmological constant, $\gamma$ and $N$ are integration constants; the parameter $N$, in particular, determines the character of the solution (existence of an event horizon, etc.). If $M>0$, $\Lambda < 0$ and $2/3<N<2$, the solution corresponds to a black hole; in this case, the asymptotic behavior is that of Anti-de Sitter spacetime. The particular case with $N=1$ corresponds to the metric of Mandal, Sengupta and Wadia (MSW) \cite{MSW}.

A recently introduced spherically symmetric solution in $2+1$ spacetime dimensions is the Schmidt--Singleton metric of the form \cite{schsin}
\be
ds^2=-Kr^2dt^2+dr^2+r^2d\theta^2,
\ee
with $K$ a constant. This solution results from a matter source in the form of a real self-interacting scalar field, which has a logarithmic behavior with the radial coordinate. The most interesting aspect of this geometry is that while its spatial part is flat, the temporal part has a behavior which corresponds to the asymptotics of Anti-de Sitter. This unusual difference between the spatial and temporal parts is associated to  the fact that the spacetime curvature is determined only by the radial pressure of the scalar field. 

In this work we mathematically introduce the shells which support traversable Lorentzian wormholes by applying the well known cut and paste procedure; then we present a general formalism for the study of the mechanical stability under perturbations preserving the symmetry of the circular shells. We apply the formalism to the geometries described above.  Finally, we discuss the results obtained. We set the units so that $G=c=1$.  

\section{General formalism}

The mathematical construction of a symmetric wormhole geometry starts from two equal copies of a  $2+1$ dimensional manifold, which in coordinates $x^{\alpha } =(t, r, \theta )$, has a metric with the generic form
\be
ds^2= -f(r)dt^2+g(r)dr^2+h(r)d\theta^2\ ,\label{metric1}
\ee
where the metric functions $f$, $g$ and $h$ are non negative from a given radius to infinity. These copies are cut and pasted at a radius $a$; when in the original manifold there is an event horizon with radius $r_h$, we take $a>r_h$. Provided that the flare-out condition is fulfilled, so that the geodesics open up at the throat, the resulting construction is a traversable wormhole. At the radius $a$ where the two copies of the geometry are joined, the components of the extrinsic curvature tensor read
\be 
K_{ij}^{\pm }=-n_{\gamma }^{\pm } \left( \frac{\partial ^{2}x^{\gamma
} } {\partial \xi ^{i}\partial \xi ^{j}}+\Gamma _{\alpha \beta }^{\gamma }
\frac{ \partial x^{\alpha }}{\partial \xi ^{i}}\frac{\partial x^{\beta }}{
\partial \xi ^{j}}\right) ,  \label{extcur}
\ee
where $\xi^i = (\tau, \theta)$ represent the coordinates on the shell, and $n_{\gamma }^{\pm }$ are the normal unit ($n^{\gamma }n_{\gamma }=1$) vectors
\be 
n_{\gamma }^{\pm }=\pm  \left| g^{\alpha \beta }\frac{\partial F}{\partial
x^{\alpha }}\frac{\partial F}{\partial x^{\beta }}\right| ^{-1/2}
\frac{\partial F}{\partial x^{\gamma }},  \label{normgen}
\ee
with $F(r)=r-a(\tau)$. Using the metric functions, the non zero components of $n_{\gamma }^{\pm }$ take the form 
\be
n_t=\mp\dot a\sqrt{g(a)f(a)},
\ee
\be 
n_r=\pm\sqrt{g(a)[1+{\dot a}^2g(a)]}.
\ee
Then, the extrinsic curvature is given by
\be
K^\pm_{\hat\tau\hat\tau}=\mp\frac{\sqrt{g(a)}}{2\sqrt{1+{\dot a}^2g(a)}}\left\{ 2\ddot a+{\dot a}^2\left[\frac{f'(a)}{f(a)}+ \frac{g'(a)}{g(a)}\right]+\frac{f'(a)}{f(a)g(a)} \right\},
\ee
\be
K^\pm_{\hat\theta\hat\theta}=\pm\frac{h'(a)}{2h(a)}\sqrt{\frac{1+{\dot a}^2g(a)}{g(a)}},
\ee
where the hats are used to denote that we are working in an orthonormal basis; the dot means a derivative with respect to the proper time $\tau $ on the shell, and a prime stands for  a derivative with respect to $r$. With these definitions we can write down the Lanczos equations \cite{daris,mus}, which relate the extrinsic curvature at both sides of the one dimensional surface with the energy-momentum tensor $S_{\hat i\hat j}=\rm{diag}(\lambda,p)$ on it
\be
-[K_{\hat i\hat j}]+[K]g_{\hat i\hat j}=8\pi S_{\hat i\hat j}.
\ee
Here the brackets denote the jump of a given quantity across the surface, $K$ is the trace of $K_{\hat i\hat j}$, and $\lambda$ and $p$ are the energy density and the pressure on the shell. The cut and paste procedure removes the interior regions $r<a$ and joins the exterior parts of the two identical geometries described by a metric like (\ref{metric1}). The jump of the extrinsic curvature components at the surface $r=a$ is associated with the linear energy density
\be
\lambda=-\frac{1}{8\pi}\frac{h'(a)}{h(a)}\sqrt{\frac{1+{\dot a}^2g(a)}{g(a)}} \label{ed}
\ee
and to the pressure
\be
p=\frac{1}{8\pi}\sqrt{\frac{g(a)}{1+{\dot a}^2g(a)}}\left\{ 2\ddot a +{\dot a}^2\left[\frac{f'(a)}{f(a)}+ \frac{g'(a)}{g(a)}\right]+\frac{f'(a)}{f(a)g(a)}\right\}.\label{pre}
\ee
It is useful to introduce the conservation equation \cite{mus}, which is obtained using the ``ADM'' constrain (also called Codazzi-Mainardi equation) and taking into account the Lanczos equations
\be
-\nabla_i S^i_j = \left[T_{\alpha \beta} \frac{\partial x^\alpha}{\partial \xi^j}
  n^\beta \right],
\label{ceq}
\ee
where the operator $\nabla$ stands for the covariant derivative and $T_{\alpha \beta}$ denotes the bulk energy-momentum tensor. Defining the one dimensional area ${\cal A}=2\pi\sqrt{h(a)}$, we can write the conservation equation in a form that relates the energy on the shell with the work done by the pressure and the energy flux
\be
\frac{d}{d\tau}(\lambda{\cal A})+p\frac{d{\cal A}}{d\tau}= -\frac{\dot a \lambda {\cal A}}{2}\left[\frac{f'(a)}{f(a)}+ \frac{g'(a)}{g(a)}+\frac{h'(a)}{h(a)}-\frac{2h''(a)}{h'(a)}\right].\label{cons}
\ee  
In the case where $g(r)=[f(r)]^{-1}$ and $h(r)=r^2$ in a neighborhood of $r=a$, it is easy to see that the factor between the brackets vanishes and the flux term is zero. By using that $\lambda'=\dot \lambda/\dot a$, we can write the condition (\ref{cons}) in the form
\be
\lambda'\frac{h(a)}{h'(a)}+\frac{\lambda+p}{2}=-\frac{\lambda h(a)}{2h'(a)}\left[\frac{f'(a)}{f(a)}+ \frac{g'(a)}{g(a)}+\frac{h'(a)}{h(a)}-\frac{2h''(a)}{h'(a)}\right]. \label{cons2}
\ee
As in the case of $3+1$ dimensions (see Ref. \cite{wh4g}), if there exists an equation of state in the form $p=p(\lambda )$ or $p=p(a,\lambda )$, the expression (\ref{cons2}) is a first order differential equation that can be recast in the form $\lambda ^{\prime}= \mathcal{F}(a,\lambda )$, for which always exists a unique solution with a given initial condition, provided that $\mathcal{F}$ has continuous partial derivatives, so it can be (formally) integrated to obtain $\lambda (a)$. Then, from Eq. (\ref{ed}) we obtain the equation of motion for the shell
\be
{\dot a}^2+V(a)=0,\label{pot}
\ee
where we have defined the potential $V(a)$ by
\be
V(a)=\frac{1}{g(a)}-\left[8\pi \lambda \frac{h(a)}{h'(a)}\right]^2.\label{potV}
\ee
The first and the second derivatives of the potential, using Eq. (\ref{cons2}) successively, have the form
\be
V'(a)=\left[\frac{1}{g(a)}\right]'+64\pi^2 \left[ W(-R+WT) \right] ,
\ee
and
\be
V''(a)=\left[\frac{1}{g(a)}\right]''-32\pi^2\left\{ \left(R-2W T\right) \left( R-W T\right)-2W^2T'+\lambda (\eta-1)\left[ S-2W\frac{h''(a)}{h'(a)}+W T\right] \right\} \label{v2}
\ee
where we have introduced $\eta=p'/\lambda'$, and the functions 
\be
R=\lambda-p,  \ \ \ \ \ \ \   S=\lambda+p, \ \ \ \ \ \ \  T=\frac{f'(a)}{f(a)}+ \frac{g'(a)}{g(a)}+ \frac{h'(a)}{h(a)},  \ \ \ \ \ \ \   W=\lambda\frac{h(a)}{h'(a)}.
\nonumber
\ee
In the static configurations, all the equations are evaluated at a fixed radius $a_0$; in this case we have the energy density and pressure for the static configuration
\be
\lambda_0=-\frac{1}{8\pi}\frac{h'(a_0)}{h(a_0)\sqrt{g(a_0)}},\label{la0}
\ee
\be
p_0=\frac{1}{8\pi}\frac{f'(a_0)}{f(a_0)\sqrt{g(a_0)}}.
\ee
Because the geometry must fulfill the flare-out condition that the geodesics open up at the wormhole throat, then $h'(a_0)>0$ is required, and from (\ref{la0}) the energy density is always negative, so that the matter at the throat is exotic, i.e. it does not satisfy the energy conditions. In the case of normal matter the parameter $\eta $ can be interpreted as the squared velocity of sound (then $0 \le \eta \le 1$); however, this is not necessarily the case here, because of the exoticism of the matter at the shell. The second derivative of the potential valued in $a_0$ can be written in terms of the metric functions in the form
\begin{eqnarray}
V''(a_0)&=& -\frac{f(a_0) f'(a_0) g'(a_0)+2 g(a_0) \left\{ [f'(a_0)]^2-f(a_0) f''(a_0)\right\} }{2 [f(a_0)]^2 [g(a_0)]^2} \nonumber \\ 
& & + \eta \frac{h(a_0) \left[ 2 g(a_0) h''(a_0)-g'(a_0) h'(a_0)\right] -2 g(a_0) [h'(a_0)]^2}{2 [g(a_0)]^2 [h(a_0)]^2}. \label{v2s} 
\end{eqnarray}
From Eq. (\ref{pot}) it is clear that a stable static solution with throat radius $a_0$ satisfies $V(a_0)=V'(a_0)=0$, and $V''(a_0)>0$. Thus the stability analysis of the wormhole configurations is essentially the analysis of the sign of the second derivative of the potential.  

\section{Examples}

In this Section we will apply the formalism to some relevant examples.  The first and the second wormhole geometries are locally indistinguishable from the exterior regions of the associated black hole metrics, while the spatial part of the third one is flat. The three cases considered share the asymptotic Anti--de Sitter behavior of the temporal part of the metric.

\subsection{The BTZ geometry}

We first consider in our wormhole construction the BTZ metric with $f(r)$ given by Eq. (\ref{2}), $g(r)=f^{-1}(r)$ and  $h(r)=r^2$. We assume a negative cosmological constant $\Lambda$, so that the asymptotic behavior of the geometry is that of the $2+1$ dimensional Anti--de Sitter universe ($AdS_3$). Wormholes connecting two copies of the exterior region (i.e. radii beyond the horizon) of this geometry have already been studied in Ref. \cite{raha}. We will revisit the construction and the linearized stability analysis of this geometry using our general formalism  in order to make straightforward the comparison with the new results presented below for wormholes associated to other spherically symmetric $2+1$ geometries. 

In the wormhole geometry resulting from the cut and paste procedure, the features of the  original metric, i.e. its horizon structure, determine the form of the stability regions in parameter space as the charge increases. Then that structure must be detailed. The position $r_h$ of the event horizon is given by the largest real positive solution of the equation $f(r)=0$, which gives $r_h=\sqrt{-M/\Lambda}$ when $Q=0$, and it can be numerically solved for $Q \neq 0$. For low values of the mass, i.e. $M<-\Lambda  r_0^2$, two critical values of the charge $Q$ exist: $Q_c^{i}$ and $Q_c^{ii}$ are such that for a charge smaller than $Q_c^{i}$ and for a charge larger than $Q_c^{ii}$ an event horizon exists in the original metric, while for $0<Q_c^{i}<|Q|<Q_c^{ii}$ there is a naked singularity. For $M=-\Lambda  r_0^2$, there is only one critical value of the charge, $Q_c^{i}=Q_c^{ii}=\sqrt{-\Lambda}r_0 $, and beyond that value, i.e. $M>-\Lambda  r_0^2$, an event horizon always exists in the original metric for any value of the charge. The critical values of the charge are obtained as the positive real solutions of the equation $Q^2 - Q^2 \ln \left[ -Q^2/(\Lambda r_0^2) \right] -M=0$. In the dynamic case, the energy density and the pressure are given respectively by Eqs. (\ref{ed}) and (\ref{pre}), with the corresponding metric functions replaced. Because in this case $g(r)=f(r)^{-1}$ and $h(r)=r^2$, the conservation equation (\ref{cons}) simplifies to
\be
\frac{d}{d\tau}(\lambda{\cal A})+p\frac{d{\cal A}}{d\tau}=0,
\ee
which gives the condition
\be
a\lambda '+\lambda+p=0.
\ee
The equation of motion of the shell (\ref{pot}) is obtained in terms of the potential by replacing the metric functions in Eq. (\ref{potV}). 
 
\begin{figure}[t!]
\centering
\includegraphics[width=16cm]{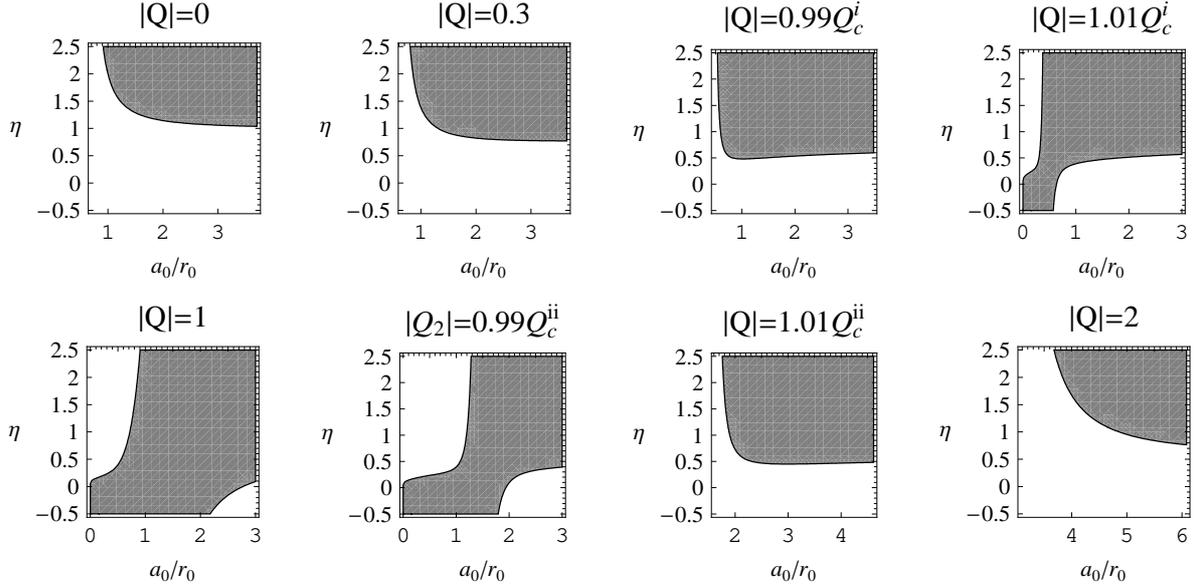}
\caption{BTZ wormhole geometry: stability regions (in gray) when two critical values of the charge exist. The values of the parameters are $\Lambda r_0^2=-1$, $M=0.5$, so that $Q_c^{i}=0.432$ and $Q_c^{ii}=1.468$.}
\end{figure}

\begin{figure}[t!]
\centering
\includegraphics[width=12cm]{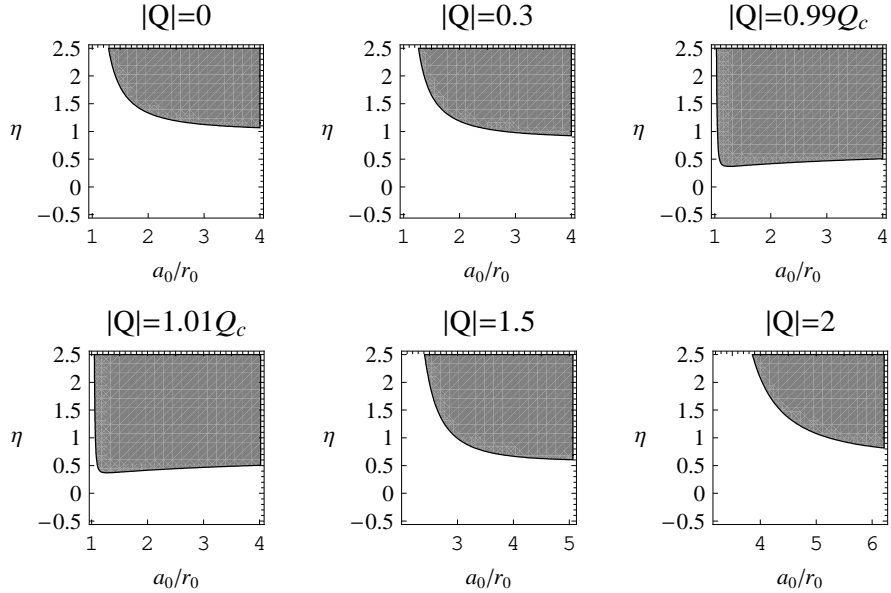}
\caption{BTZ wormhole geometry: stability regions (in gray) for the case in which there is only one value of the critical charge. The values of the parameters are $\Lambda r_0^2=-1$, $M=1$,  so that $Q_c^i=Q_c^{ii}=1$.}
\end{figure}

\begin{figure}[t!]
\centering
\includegraphics[width=16cm]{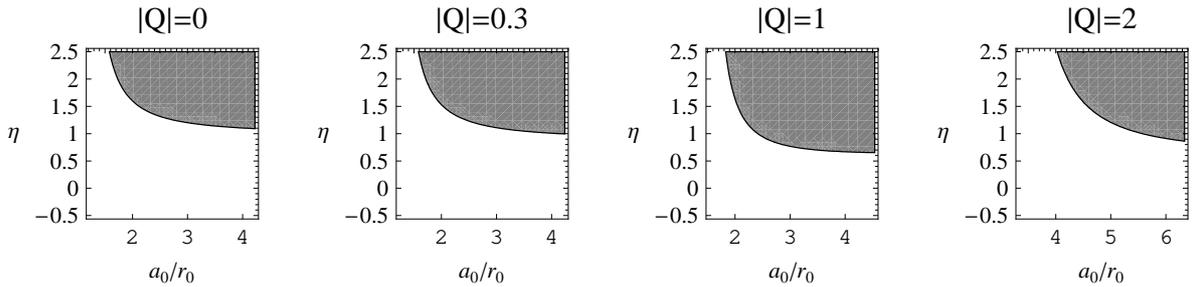}
\caption{BTZ wormhole geometry: stability regions (in gray) when there is no critical charge, so an event horizon always exists in the original manifold. The values of the parameters are $\Lambda r_0^2=-1$, $M=1.5$.}
\end{figure}

The energy density and the pressure in the static case, with throat radius $a_0$, take the form
\be
\lambda _0=-\frac{\sqrt{-M-\Lambda a_0^2 -2 Q^2 \ln (a_0/r_0)}}{4 \pi  a_0}
\ee
and
\be
p_0=-\frac{\Lambda a_0^2 + Q^2}{4 \pi  a_0 \sqrt{-M -\Lambda a_0^2 -2 Q^2 \ln (a_0/r_0)}}.
\ee
The second derivative of the potential, evaluated at $a_0$, reads
\be
V''(a_0)=\frac{2}{a_0^2} \left[ -\Lambda a_0^2 +Q^2+ \frac{\left( \Lambda a_0^2 +Q^2 \right)^2 }{M+ \Lambda a_0^2  +2 Q^2 \ln (a_0/r_0)}+ \left[ M-  Q^2 + 2 Q^2 \ln (a_0/r_0) \right] \eta \right] .
\ee

Figures 1 to 3 illustrate the behavior of the stability regions $(V''(a_0)>0)$ with an increase of the charge, for a fixed value of the adimensionalized cosmological constant $\Lambda r_0^2$ and three different values of the mass $M$. In particular, only in the case in which there is no  horizon in the original metric, stability could be possible for vanishing or small negative $\eta$. In all cases a null charge makes stability incompatible with $|\eta|<1$. The evolution of the regions is not monotonous: as a larger charge is considered, so that the original geometry has an event horizon, the stability regions recover a form similar to those corresponding to low values of the charge. The analysis here extends the previous works \cite{raha} to a larger range of the relevant parameters.

\subsection{The Chan-Mann metric}

In the  case of the symmetric wormhole connecting two charged Chan-Mann solutions, having a cosmological constant and a dilaton field, the metric functions correspond to $f(r)$ given by Eq. (\ref{cm}), $g(r)=4r^{\frac{4}{N}-2}{N^{-2}\gamma^{-\frac{4}{N}}f^{-1}(r)}$ and $h(r)=r^2$.  The associated scalar field is 
\be 
\phi=\frac{2k}{N}\ln \left( \frac{r}{\beta }\right) ,
\ee
with $\beta=\gamma^{2/(2-N)}$ and $k=\pm\sqrt{N(2-N)(2B)^{-1}}$ ($B$ is a constant). When $2/3<N<2$ the solution represents a black hole if $0\leq |Q|\leq Q_c$ or a naked singularity if $|Q|> Q_c$, where the critical value of the charge is given by
\be
Q_c=\sqrt{-\Lambda }  \left[\frac{(N-2) (3 N-2) M}{8 N \Lambda }\right]^{N/(3N-2)}.
\ee
When $0\leq |Q|\leq Q_c$, for the wormhole construction we take two copies of the region $r\geq a> r_h$, where $r_h$ is the horizon radius determined by the greatest positive root of the function $f(r)$. When $|Q|>Q_c$ there is no horizon and then no restriction exists on the possible radius of the throat. The energy density and the pressure for the dynamic case are obtained by replacing the corresponding metric functions in Eqs. (\ref{ed}) and (\ref{pre}). In this case, Eqs. (\ref{cons}) and (\ref{cons2}) have, respectively, the form
\be
\frac{d}{d\tau}(\lambda{\cal A})+p\frac{d{\cal A}}{d\tau}=-2\pi\dot a\lambda\left(\frac{2}{N}-1 \right)
\ee
and
\be
a\lambda'+\lambda+p=-\lambda\left(\frac{2}{N}-1 \right).
\ee
One obtains the equation of motion of the throat by introducing the metric functions in Eq. (\ref{potV}) and replacing this potential in Eq. (\ref{pot}).

\begin{figure}[t!]
\centering
\includegraphics[width=12cm]{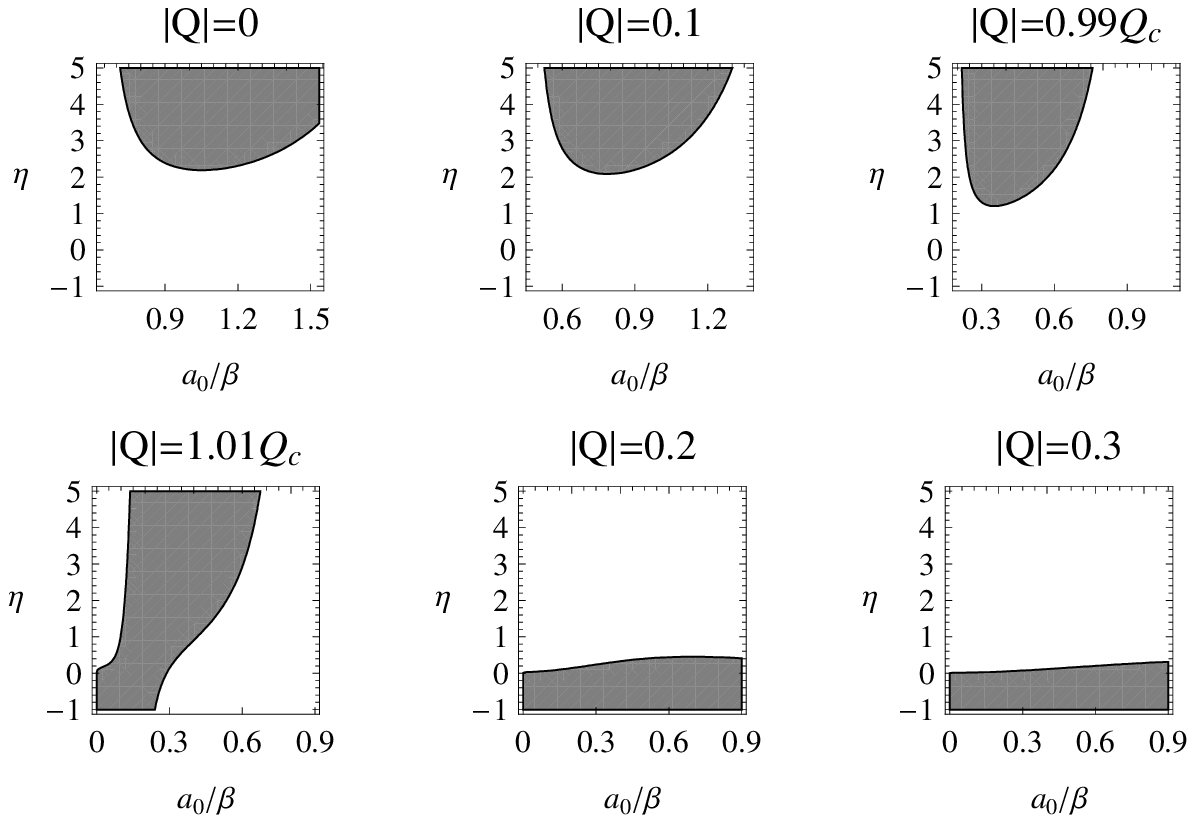}
\caption{Chan-Mann wormhole geometry: stability regions  (in gray) for the case $N=9/5$, $\Lambda\beta^2= -1$, $M\beta^{(2-N)/N}=0.5$, for which $Q_c=0.137$.}
\end{figure}
\begin{figure}[t!]
\centering
\includegraphics[width=12cm]{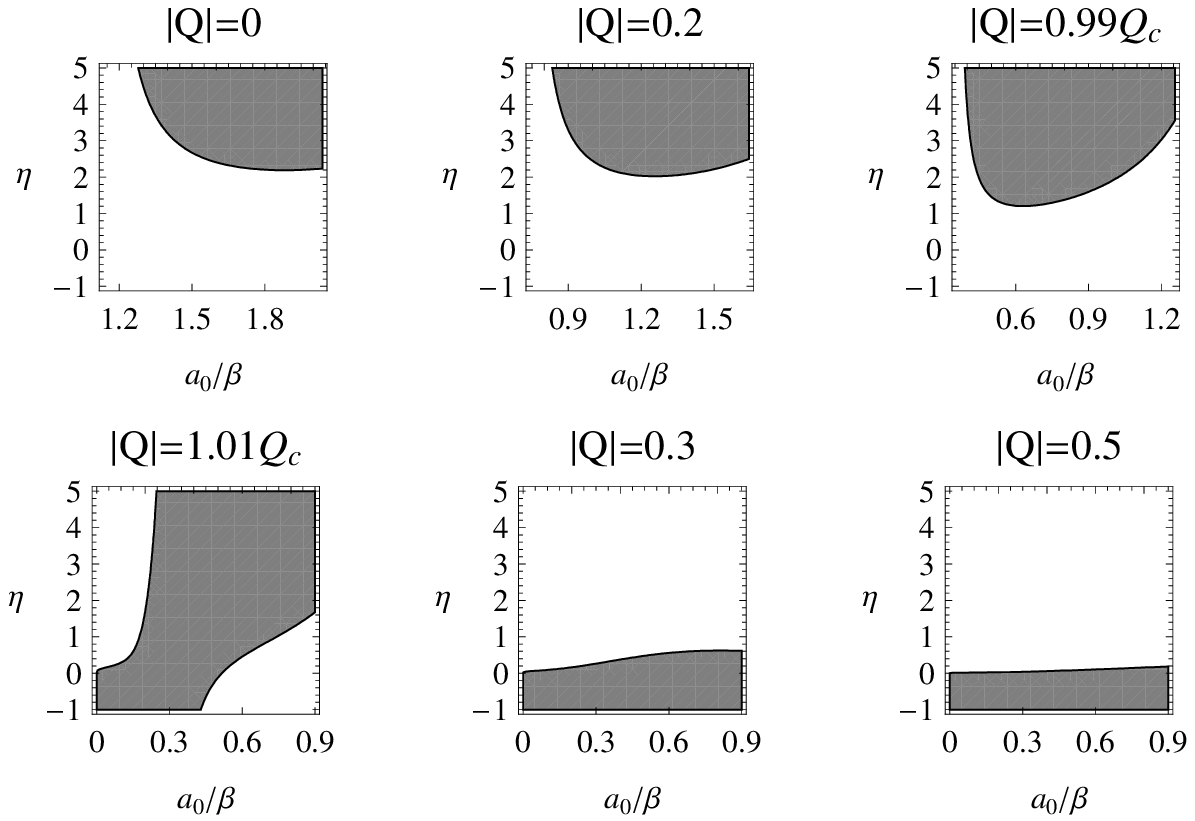}
\caption{Chan-Mann wormhole geometry: stability regions  (in gray) for the case  $N=9/5$, $\Lambda\beta^2= -1$, $M\beta^{(2-N)/N}=1.5$, for which $Q_c=0.246$.}
\end{figure}

\begin{figure}[t!]
\centering
\includegraphics[width=12cm]{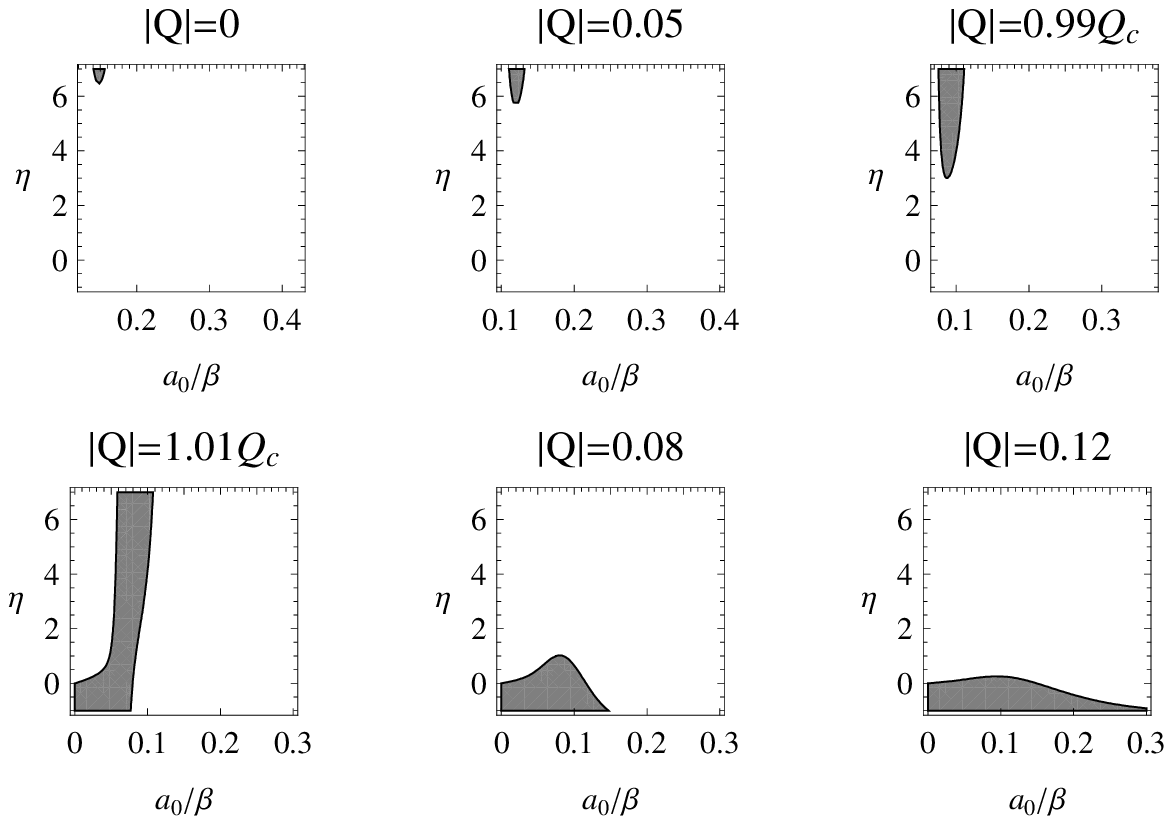}
\caption{Chan-Mann wormhole geometry: stability regions  (in gray) for the case  $N=1$ (MSW metric), $\Lambda\beta^2= -1$, $M\beta^{(2-N)/N}=0.5$, for which $Q_c=0.0625$.}
\end{figure}
\begin{figure}[t!]
\centering
\includegraphics[width=12cm]{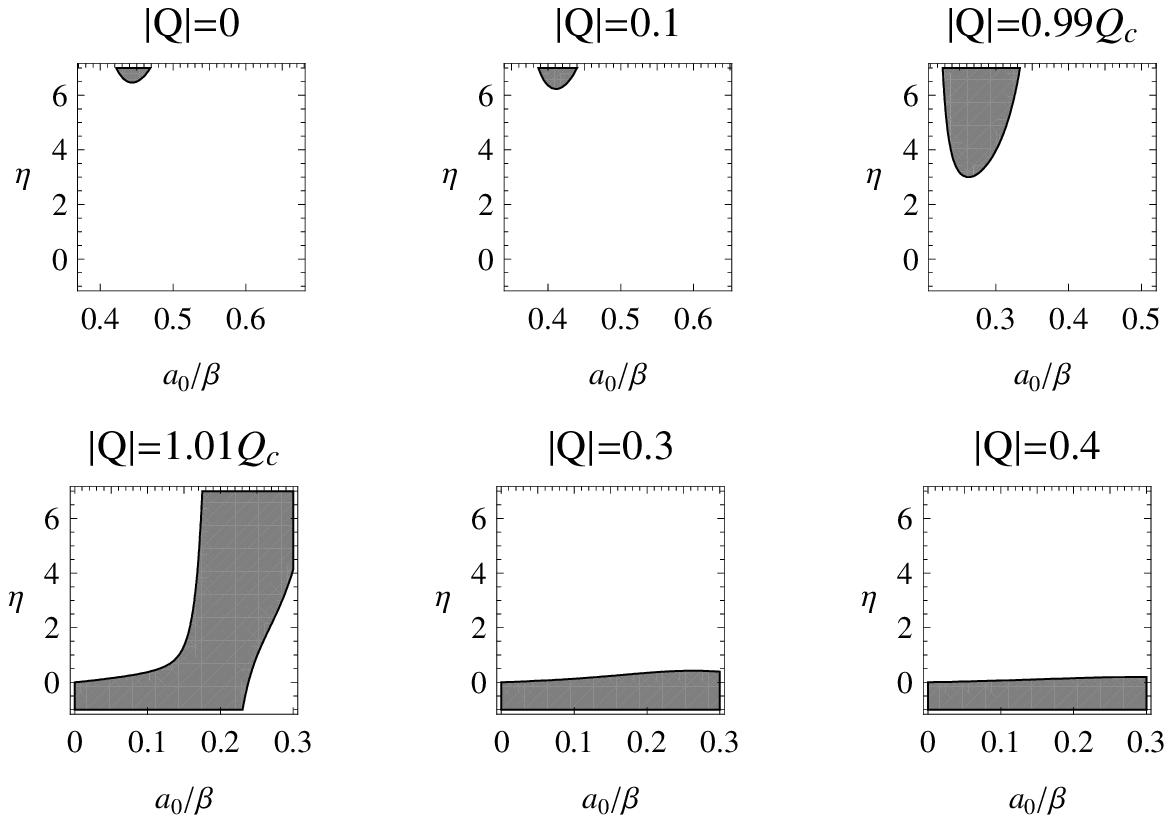}
\caption{Chan-Mann wormhole geometry: stability regions  (in gray) for the case  $N=1$ (MSW metric), $\Lambda\beta^2= -1$, $M\beta^{(2-N)/N}=1.5$, for which $Q_c=0.1875$.}
\end{figure}
 
The static configurations with shell radius $a_0$, have the energy density and the pressure given by
\be
\lambda_0=-\frac{\gamma ^{2/N} \sqrt{N} }{4 \pi a_0^{2/N} } \sqrt{-\frac{ M a_0^{-1+2/N}}{2}-\frac{2 \Lambda  a_0^2}{3 N-2}+\frac{2 Q^2}{2-N}}
\ee
and
\be
p_0=\frac{\gamma ^{2/N}\sqrt{2-N} } {8 \pi a_0^{2/N} \sqrt{2N (3 N-2)}} \frac{ M (2-3 N) (2-N) a_0^{-1+2/N}-8 \Lambda  N a_0^2 } {\sqrt{M (2-3 N) (2-N) a_0^{-1+2/N}-4  \Lambda  (2-N) a_0^2 +4 (3 N-2) Q^2}}.
\ee
The second derivative of the potential evaluated at $a_0$ has the form
\be
V''(a_0)=\left( \psi _0 +\chi _0 \eta \right) a_0^{-(4+N)/N}\gamma ^{4/N},
\ee
where
\begin{eqnarray}
\psi _0 &=& \frac{(2-N) \left[ M (N+2) a_0^{2/N}+4 a_0 Q^2\right] }{4 N}-\frac{a_0^3 \Lambda  \left(13 N^2-20 N+4\right)}{N (3 N-2)}\nonumber \\
& & + \frac{4 a_0^2 (N-2) (3 N-2) \left(a_0^2 \Lambda +Q^2\right)^2}{N \left[ M (N-2) (3 N-2) a_0^{2/N}+4 a_0^3 \Lambda  (N-2)+4 a_0 (3 N-2) Q^2\right] }
\end{eqnarray}
and
\be
\chi_0=\frac{M (N+2) a_0^{2/N}}{2}-\frac{4 a_0 \left[ a_0^2 \Lambda (N-2)^2+(4-6 N) Q^2\right] }{(N-2) (3 N-2)}.
\ee

The stability results (which correspond to $(V''(a_0)>0$) are shown for some representative values of the parameters in Figs. 4 to 7. For fixed values of the parameters $N$, the adimensionalized cosmological constant $\Lambda \beta^2$, and the adimensionalized mass $M\beta^{(2-N)/N}$, the largest ranges of the parameter $\eta$ compatible with stability take place for values of the charge near the critical one; in particular, large ranges of $\eta$ including values within the interval $[0,1)$ correspond to stable solutions if the charge is slightly above the critical value. In general, stable configurations with $0\leq\eta <1$ require $|Q|>Q_c$. The behavior with $M\beta^{(2-N)/N}$ for a fixed parameter $N$ and $\Lambda \beta^2$ only shows a change in the range of $a_0/\beta$ for which configurations are stable, without changing the range of the parameter $\eta$. For given  $\Lambda \beta^2$ and $M\beta^{(2-N)/N}$, decreasing values of $N$ (within the range $2/3<N<2$) move the stability regions to smaller $a_0/\beta$, and for a fixed $|Q|<Q_c$ to higher positive values of the parameter $\eta$.

\subsection{The Schmidt--Singleton metric}

Now we consider the particular case of a wormhole connecting two exterior geometries with the Schmidt--Singleton metric given by $f(r)=Kr^2,\  g(r)=1,\  h(r)=r^2$. The associated scalar field is given by
\be 
\phi=\frac{1}{\sqrt{\kappa}}\ln \left( \frac{r}{r_0}\right) ,
\ee
where $\kappa$ is the coupling constant in the Liouville potential of the field and $r_0$ is a constant. The metric presents no event horizon, so there is no restriction to the possible radius for the wormhole throat. After applying the formalism, we have the energy density and pressure given by the simple expressions
\be
\lambda=-\frac{1}{4\pi a}\sqrt{1+{\dot a}^2},
\ee
\be
p = \frac{1}{4\pi}\frac{1}{\sqrt{1+{\dot a}^2}}\left[\ddot a+\frac{1+{\dot a}^2}{a} \right].
\ee
The conservation equation for this case is
\be
\frac{d}{d\tau}(\lambda{\cal A})+p\frac{d{\cal A}}{d\tau}=-2\pi\dot a \lambda 
\label{sce1}
\ee
which gives the condition
\be
\lambda' a+2\lambda+p=0.
\label{sce2}
\ee
The resulting equation of motion is ${\dot a}^2+V(a)=0$ where the potential has the form
\be
V(a)= 1-\left( 4\pi a \lambda \right)^2.
\ee
The simplicity of $V(a)$ makes possible a fully analytical treatment; no plots are needed in order to obtain the conditions required to render the wormhole construction stable under radial perturbations. Using Eq. (\ref{sce2}), the first and second derivatives of $V(a)$ read
\be
V'(a)=32\pi^2 a \lambda \left( \lambda + p \right),
\ee
\be
V''(a)=-32\pi^2\left\{\lambda^2+(2\lambda+p)\left[(\eta+1)\lambda + p\right] \right\}.
\ee
In the static case the  energy density and pressure are
\be
\lambda_0=-\frac{1}{4\pi a_0},\ \ \ \ \ \ \ \ \ p_0=\frac{1}{4\pi a_0}.
\ee
Hence, the second derivative of the potential evaluated at the radius of the static configuration is given by
\be
V''(a_0)=-2\frac{\left(\eta+1\right)}{a_0^2}.
\ee
This result implies that the mechanical stability under radial perturbations is possible only for the interval $\eta<-1$. Had the ring been constituted by normal matter, the requirement of negative values of the parameter $\eta$ would rule out stability. But the matter of the ring does not fulfill the energy conditions, i.e. we are dealing with an exotic fluid, then the result $\eta< -1$ does not necessarily discard stable configurations. In particular, phantom energy, common in current cosmology, has an equation of state with this feature.

\section{Discussion}

We have developed a general formalism for the construction and the analysis of stability of  the static configurations corresponding to circular thin-shell wormholes in $2+1$ spacetime dimensions. We have used the usual cut and paste procedure in the construction, and we have adopted a linearized equation of state for the exotic matter at the throat. The stability analysis then has been reduced to the study of the sign of the second derivative of an effective potential evaluated at the throat radius $a_0$. We have applied this formalism to three examples: the charged BTZ, Chan--Mann, and Schmidt--Singleton geometries. Though the steps followed are the same, the difficulties are different in each case. In particular, in two examples the energy flux in the right hand side of the energy conservation equation does not vanish, while in the other one (BTZ) it is identically zero. 

In all cases we have found the stability regions in terms of the adimensionalized throat radius and the parameter $\eta$ associated with the equation of state. In the BTZ wormhole, for low values of the mass and when the charge is very close or between the two critical values (corresponding to the extremal black hole in the original metric) small values of $\eta$, in particular with $0\leq \eta < 1$, are compatible with stability. If the mass is higher than a certain value, this feature is lost. The results corresponding to the Chan--Mann wormhole are qualitatively similar, when the charge is very close to the critical one, to those of the low mass BTZ geometry. The main difference is that the behavior does not change for large values of the mass as in the BTZ case. The Chan--Mann metric is characterized by the parameter $N$ (with $2/3<N<2$). Decreasing values of $N$ shift the regions of stability to smaller adimensionalized throat radius, and to higher positive values of the parameter $\eta$ when the charge is lower than the critical one. The Schmidt--Singleton wormhole spacetime, which has no charge and admits the simplest treatment, is stable  only when $\eta< -1$. This characteristic makes this geometry less interesting than the other two analyzed in our work, but does not rule out its stability.

\section*{Acknowledgments}

This work has been supported by Universidad de Buenos Aires and CONICET.

\end{document}